\documentclass[showpacs,showkeys]{revtex4}
\usepackage{graphicx}
\pdfoutput=1
\def\lsim{\mathrel{\rlap{\lower4pt\hbox{\hskip1pt$\sim$}}
    \raise1pt\hbox{$<$}}}         
\def\gsim{\mathrel{\rlap{\lower4pt\hbox{\hskip1pt$\sim$}}
    \raise1pt\hbox{$>$}}}         

\begin{document}

\title{Nuclear Constraints on the Weak \\
Nucleon-Nucleon Interaction}

\author{W. C. Haxton}

\affiliation{Institute for Nuclear Theory and Department of Physics, \\
University of Washington, Seattle, WA 98195, USA \\
Email: haxton@phys.washington.edu}

\begin{abstract}
I discuss the current status of efforts to constrain the strangeness-conserving
weak hadronic interaction, which can be isolated in nuclear systems because
of the associated parity violation.  
\end{abstract}


\maketitle

\section{Parity Nonconservation in the NN System}
In this talk I will discuss the weak nucleon-nucleon (NN) interaction: the experiments 
that have been done, the strategies theorists have developed to interpret
measurements, and the puzzles that remain to be resolved.  

While the weak interaction can be observed in
flavor-changing hadronic decays, the neutral current contribution to such
decays is suppressed by the GIM mechanism and thus unobservable.
The NN and nuclear systems are thus the only practical laboratories for studying
the hadronic weak interaction in all of its aspects \cite{adelberger,cp}.

As the weak contribution to the NN interaction is many orders of magnitude
smaller than the strong and electromagnetic contributions (the suppression
relative to the strong interaction is $\sim 4 \pi G m_\pi^2/g^2_{\pi NN} \sim 10^{-7}$), parity
violation must be exploited to isolate weak observables.   The most common
observables are pseudoscalars arising from the interference of the weak and
strong/electromagnetic interactions, e.g., the circular polarization of $\gamma$ rays emitted
from an unpolarized excited nuclear state, or the $\gamma$ ray asymmetry 
if the nuclear state can be polarized.  The observable depends on an interference
between parity-conserving (PC) and parity-non-conserving (PNC) amplitudes,
and the weak interaction appears linearly.  Alternatively, there are processes, such as
the $\alpha$ decay of an unnatural-parity nuclear state to a 0$^+$ final
state, where the amplitude is entirely weak.  Such observables are proportional
to the squares of weak matrix elements, and thus
are not associated with a pseudoscalar.

The range of the underlying weak interaction, mediated by W and Z exchange, is 
$\sim$ 0.002 fm, much smaller than the radius of the nucleon.  For this reason
the nuclear weak force is often modeled as a series of meson exchanges,
with one nucleon vertex strong and with the second vertex containing the
weak physics, as depicted in Fig. \ref{fig:meson}.  The resulting isospin of the
weak meson-nucleon coupling is related to the underlying currents in an interesting way.
The hadronic weak interaction has the low-energy current-current form
\begin{equation}
L^{eff} = {G \over \sqrt{2}} \left[ J^\dagger_W J_W + J_Z^\dagger J_Z \right] + h.c.
\end{equation}
where the charge-changing current is the sum of $\Delta$I=1 $\Delta$S=0 and Cabibbo-suppressed
$\Delta$I=1/2 $\Delta$S=-1 terms,
\begin{equation}
J_W = \cos{\theta_C} J_W^{\Delta S=0} + \sin{\theta_C} J_W^{\Delta S =-1}.
\end{equation}
Consequently the $\Delta$S=0 interaction has the form
\begin{equation}
L^{eff}_{\Delta S=0} = {G \over \sqrt{2}} \left[ \cos^2{\theta_C}J_W^{0 \dagger} J_W^0 +
\sin^2{\theta_C} J_W^{1 \dagger} J_W^1+ J_Z^\dagger J_Z \right]
\end{equation}
where the first term, a symmetric product of $\Delta$I=1 currents, has $\Delta$ I=0,2, while the
second term, a symmetric product of $\Delta$ I=1/2 currents, is $\Delta$ I=1 but
Cabibbo suppressed.  Consequently a $\Delta$I=1 PNC meson-nucleon vertex should
be dominated by the neutral current term -- a term not accessible in strangeness-changing
processes.  One could isolate this term by an isospin analysis of a complete
set of PNC NN observables.

\begin{figure}
\includegraphics[width=15cm]{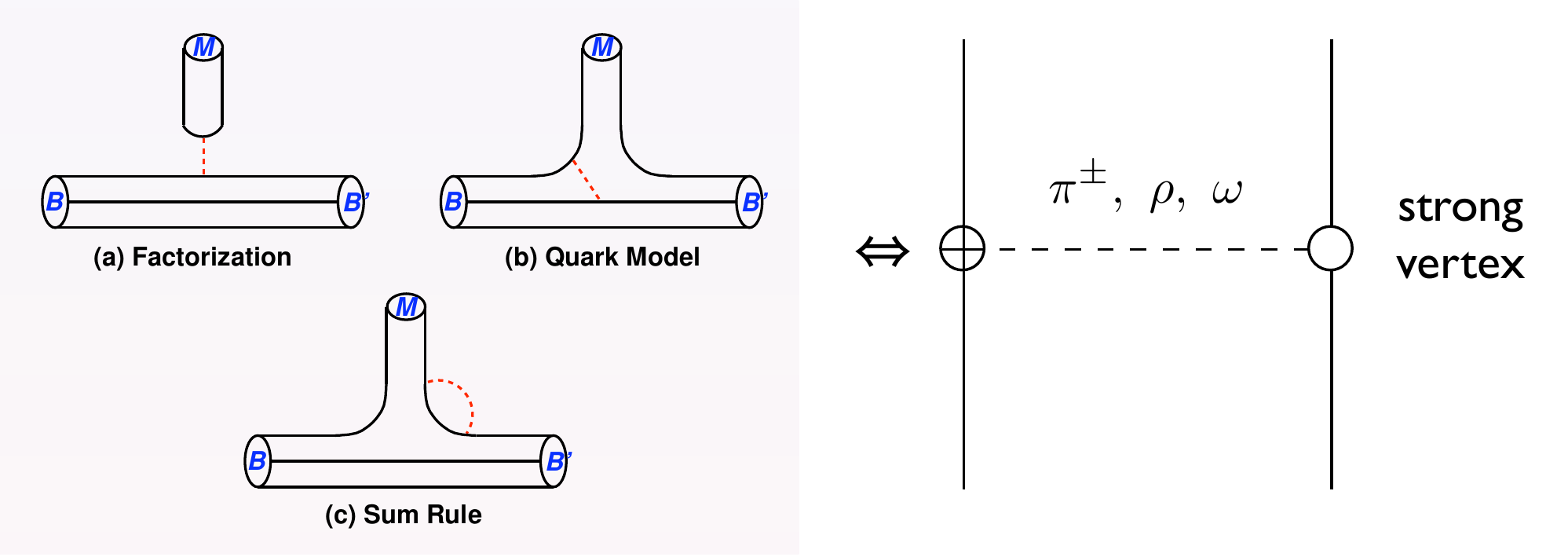}
\caption{A single-boson-exchange contribution to $V_{PNC}$ contains one weak vertex (left) and one strong one (right).  The weak vertex is decomposed into the quark-level terms that DDH estimated, using the standard model in combination with techniques such as factorization, the quark model, and sum rules.}
\label{fig:meson.pdf}
\end{figure}

\section{S-P Amplitudes and Meson-Exchange Potentials}
There are several ways to describe low-energy PNC NN interactions.  Perhaps the
simplest representation, the Danilov amplitudes, is an S-P partial wave description appropriate in the 
low-momentum limit.  Table \ref{tab:one} lists the five partial waves.  The coefficients multiplying
these amplitudes can be treated as free
parameters, to be determined from experiments.  Once these are fixed, other
low-energy PNC observables could be predicted, in virtually a model-independent
way.

\begin{table}
\caption{S-P weak PNC amplitudes and the corresponding meson-exchanges \cite{adelberger}.}
\begin{tabular}{@{}lcccccc@{}}
\toprule
Transition & I $\leftrightarrow$ I$^\prime$ & $\Delta$ I & n-n & n-p & p-p & meson exchanges \\\colrule
${}^3S_1 \leftrightarrow {}^1P_1$ & 0 $\leftrightarrow$ 0 & 0 & & x & & $\rho,\omega$ \\
${}^1S_0 \leftrightarrow {}^3P_0$ & 1 $\leftrightarrow$ 1 & 0 & x & x & x & $\rho,\omega$ \\
 & & 1 & x & & x & $\rho,\omega$\\
 & & 2 & x & x & x & $\rho$ \\
 ${}^3S_1 \leftrightarrow {}^3P_1$ & 0 $\leftrightarrow$ 1 & 1 & & x & & $\pi^\pm,\rho,\omega$ \\\botrule
\end{tabular}
\label{tab:one}
\end{table}

A second approach expresses the interaction as a set of single meson exchanges
(see Fig. \ref{fig:meson}), in analogy with traditional meson-exchange 
treatments of the strong force, but with one of the strong vertices replaced by
a weak one containing the short-range W,Z physics.  The possible
exchanges are constrained by symmetries, e.g., Barton's theorem excludes on-shell
couplings to neutral scalar mesons.  If one includes $\rho$, $\omega$, and $\pi^\pm$
exchanges, one has enough freedom to reproduce the five Danilov
amplitudes and to model the long-range pion contribution important to higher partial waves.

Much of the work that has been done in the field uses a potential developed by
Donoghue, Desplanques, and Holstein (DDH) \cite{ddh}
\begin{eqnarray}
2 M V^{PNC}(\vec{r}) &=& {g_{\pi NN} f_\pi \over \sqrt{2}} \tau_\times^z \vec{\sigma}_+ \cdot \vec{u}_\pi \nonumber \\
&-& g_\rho \left[h_\rho^0 \vec{\tau}_1 \cdot \vec{\tau}_2 + h_\rho^1 \tau_+^z+ h_\rho^2 \tau^{zz}\right]
\left[(1+\mu_v) \vec{\sigma}_\times \cdot \vec{u}_\rho + \vec{\sigma}_- \cdot \vec{v}_\rho \right] \nonumber \\
&-& g_\omega \left[h_\omega^0+h_\omega^1 \tau_+^z\right] \left[(1+\mu_s) \vec{\sigma}_\times \cdot \vec{u}_\omega + \vec{\sigma}_- \cdot \vec{v}_\omega \right] \nonumber \\
&-& \tau_-^z \vec{\sigma}_+ \cdot \left[ g_\omega h_\omega^1 \vec{v}_\omega - g_\rho h_\rho^1 \vec{v}_\rho \right] - \tau_\times^z g_\rho h_\rho^{'1} \vec{\sigma}_+ \cdot \vec{u}_\rho
\end{eqnarray}
where
\begin{eqnarray}
\vec{\sigma}_\times \equiv i ~\vec{\sigma}_1 \times \vec{\sigma}_2~~~~~~\vec{\sigma}_+ \equiv {1 \over 2} \left[\vec{\sigma}_1+ \vec{\sigma}_2 \right]~~~~~~\vec{\sigma}_- \equiv \vec{\sigma}_1-\vec{\sigma}_2 ~~~~~\nonumber \\
\tau^{zz} \equiv {1 \over 2 \sqrt{6}} \left[ 3 \tau^z_1 \tau^z_2-\vec{\tau}_1 \cdot \vec{\tau}_2 \right] ~~~~~~~~~~~~~~~~~~~~ \nonumber \\
\vec{u}(\vec{r}) \equiv \left[ \vec{p},e^{-m r}/4 \pi r \right]~~~~~\vec{v}(\vec{r}) \equiv  \left\{ \vec{p},e^{-m r}/4 \pi r \right\}~~~~~ \vec{p} \equiv \vec{p}_1-\vec{p}_2.
\end{eqnarray}
The $g_{\pi NN}$, $g_\rho$, and $g_\omega$ ($f_\pi$, $h_\rho$, and $h_\omega$) are the strong (weak) $\pi^\pm$, $\rho$, and $\omega$ couplings.  As noted previously, the limit $m_\rho,m_\omega \rightarrow \infty$ maps the short-range part of this potential onto the five Danilov amplitudes.
(As there are six short-range $\rho$ and $\omega$ couplings, there is a transformation
among these couplings that leaves the S-P amplitudes unchanged \cite{adelberger}.)
The estimated parameter
ranges and best values recommended by DDH are shown in Table \ref{tab:two}, along with
several other parameterizations of this potential.  Such a meson-exchange treatment, by
providing a model for P-D and other higher partial-waves, presumably has some
validity when extended to higher momenta: analogous treatments of the strong potential
are quite successful in describing intermediate-range NN interactions.

\begin{table}
\caption{Recommended ranges and best values for the DDH potential, along with three
other parameterizations. From \cite{cp}.}
\begin{tabular}{@{}cccccc@{}}
\toprule
Coupling ($\times 10^{-7}$) & DDH Range \cite{ddh} & Best \cite{ddh} & DZ \cite{dz} & FCDH \cite{fcdh} & KM \cite{km} \\\colrule
$f_\pi$ & 0.0$\leftrightarrow$11.4 & 4.6 & 1.1 & 2.7 & 0.2 \\
$h_\rho^0$ & -30.8$\leftrightarrow$11.4 & -11.4 & -8.4 & -3.8 & -3.7  \\
$h_\rho^1$ & -0.4 $\leftrightarrow$ 0.0 & -0.2 & 0.4 & -0.4 & -0.1 \\
$h_\rho^2$ & -11.0$\leftrightarrow$ -7.6 & -9.5 & -6.8 & -6.8 & -3.3 \\
$h_\omega^0$ & -10.3 $\leftrightarrow$ 5.7 & -1.9 & -3.8 & -4.9 & -6.2 \\
$h_\omega^1$ & -1.9 $\leftrightarrow$ -0.8 & -1.1 & -2.3 & -2.3 & -1.0 \\
$h_\rho^{'1}$ & & 0.0 & & & -2.2  \\\botrule
\end{tabular}
\label{tab:two}
\end{table}
\begin{figure}
\includegraphics[width=15cm]{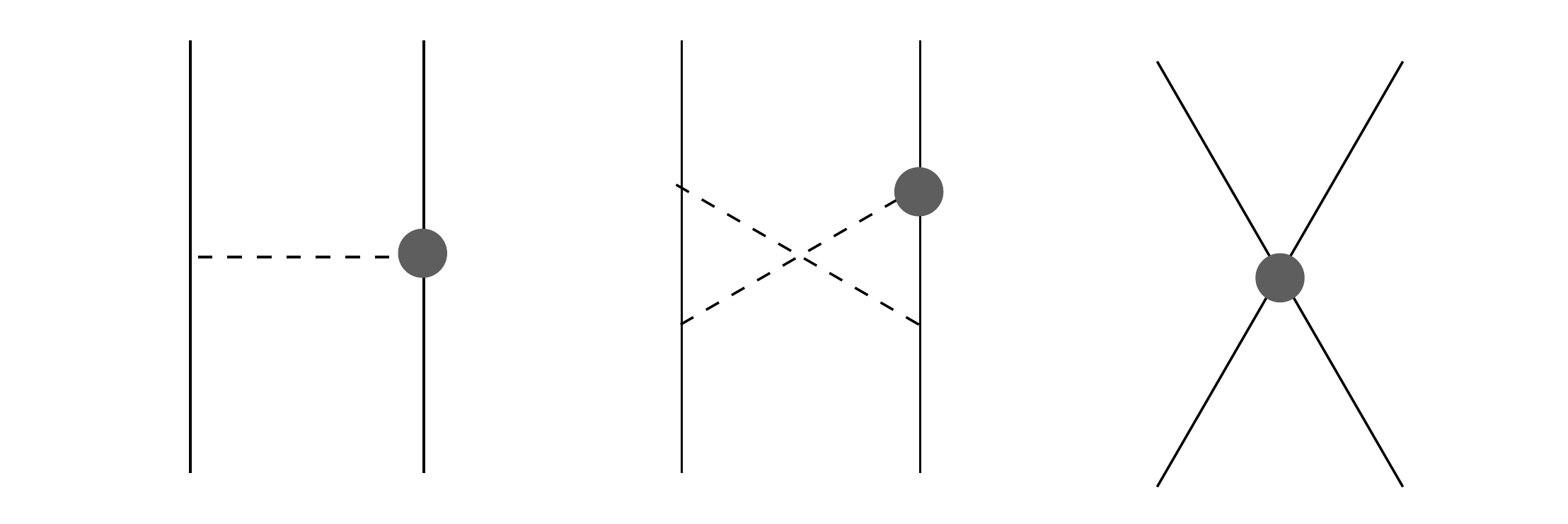}
\caption{Contributions in a chiral perturbation theory expansion of $V_{PNC}$ including long-range pion exchange, the intermediate-range contribution from crossed pions, and S-P contact
interactions \cite{zhu}.}
\label{fig:eft}
\end{figure}

A third approach, developed recently, is a fully systematic expansion in chiral perturbation
theory \cite{zhu} in terms of $m_\pi/\Lambda_{\chi SB}$.  This allows for a consistent treatment
of iterated pions, helping to define the PNC potential at intermediate ranges: the contributions,
illustrated in Fig. \ref{fig:eft}, include such terms, long-range single $\pi^\pm$ 
exchange, and five contact interactions.  While ten
parameters arise in this approach in leading order, only five are independent in the limit
of low momentum \cite{zhu,cpl}.

\section{Experimental Constraints}
The goal of the field has been to determine the weak meson-nucleon coupling strengths by fitting
to experiment.  If the nonperturbative strong interaction physics associated with the
meson-nucleon vertices can be computed, one would then be able to connect these
vertices with the underlying elementary couplings of the standard model.  Ideally
one would make a complete set of measurements in the NN system.  However, in most cases the
required sensitivity is difficult to achieve.  The longitudinal analyzing power for $\vec{p}+p$
has been measured at Bonn \cite{bonnpp} and SIN \cite{sinpp},
\begin {eqnarray}
A_L^{\vec{p}+p}(13.6~\mathrm{MeV}) &=& (-0.93 \pm 0.21) \times 10^{-7}~~(\mathrm{Bonn}) \nonumber \\
A_L^{\vec{p}+p}(45 ~\mathrm{MeV}) &=& (-1.57 \pm 0.23) \times 10^{-7}~~(\mathrm{SIN}) ,
\end{eqnarray}
constraining the $^1S_0-{}^3P_0$ $\Delta$I=0,1,2 amplitudes.  The circular polarization
of the $\gamma$s produced in $n+p$ radiative capture has also been measured \cite{pnp},
\begin{equation}
P_\gamma(n+p \rightarrow d + \gamma) = (1.8 \pm 1.8) \times 10^{-7}.
\end{equation}
$P_\gamma$ depends on the $^1S_0-{}^3P_0$ $\Delta$I=0,2 and $^3S_1-{}^1P_1$ $\Delta$I=1 amplitudes.
Finally, there is a upper bound on the $\gamma$-ray  asymmetry in radiative capture \cite{nprad}
\begin{equation}
A_\gamma(\vec{n}+p \rightarrow d + \gamma) = (0.6 \pm 2.1) \times 10^{-7}.
\end{equation}
$A_\gamma$ depends on the $^3S_1-{}^3P_1$ $\Delta$I=1 amplitude.  A program has begun
at LANSCE and will continue at the SNS to improve this result, with a factor of 20 the
long-term goal.  There are also plans to measure the spin rotation of 
polarized neutrons passing through parahydrogen at the SNS \cite{markoff}.

As there are quasi-exact methods for treating few-body nuclei, PNC observables in
such systems can also be interpreted reliably.  The analyzing power for polarized protons
scattering on $^4$He has been measured \cite{hespin}
\begin{equation}
A_L^{\vec{p}+^4\mathrm{He}}(46~\mathrm{MeV}) = (-3.3 \pm 0.9) \times 10^{-7}~~(\mathrm{SIN}).
\end{equation}
This ``odd proton" observable depends on a combination of isovector and isoscalar
couplings quite similar to that tested in $^{19}$F, discussed below.  There are
also two bounds of interest,
\begin{eqnarray}
{d \over dz} \phi^{\vec{n}+\alpha}_n(\mathrm{thermal}) &=& (8 \pm 14) \times 10^{-7}~ \mathrm{rad/m} ~~(\mathrm{NIST}~\rightarrow \mathrm{SNS}) \nonumber \\
A_L^{\vec{p}+d}(15~\mathrm{MeV}) &=& (-0.35 \pm 0.85)~~(\mathrm{LANL}).
\end{eqnarray}
The NIST effort on the neutron spin rotation is in progress \cite{mich}.  There are
plans to continue the work at the SNS.

Measurements in complex nuclei comprise the third class of experiments.  One advantage of
nuclear experiments is  the opportunity, because of level degeneracies and
favorable PNC/PC matrix element ratios, to significantly enhance
the size of PNC observables.  There are nuclear PNC effects of $\sim$ 10\%, in contrast to
the 10$^{-7}$ characteristic of the NN system.  One can also use isospin and other
nuclear quantum numbers as a filter, isolating specific components of the PNC interaction.
The disadvantage of such systems is wave function complexity, which complicates the extraction
of coupling strengths from the observables.

\begin{figure}
\includegraphics[width=15cm]{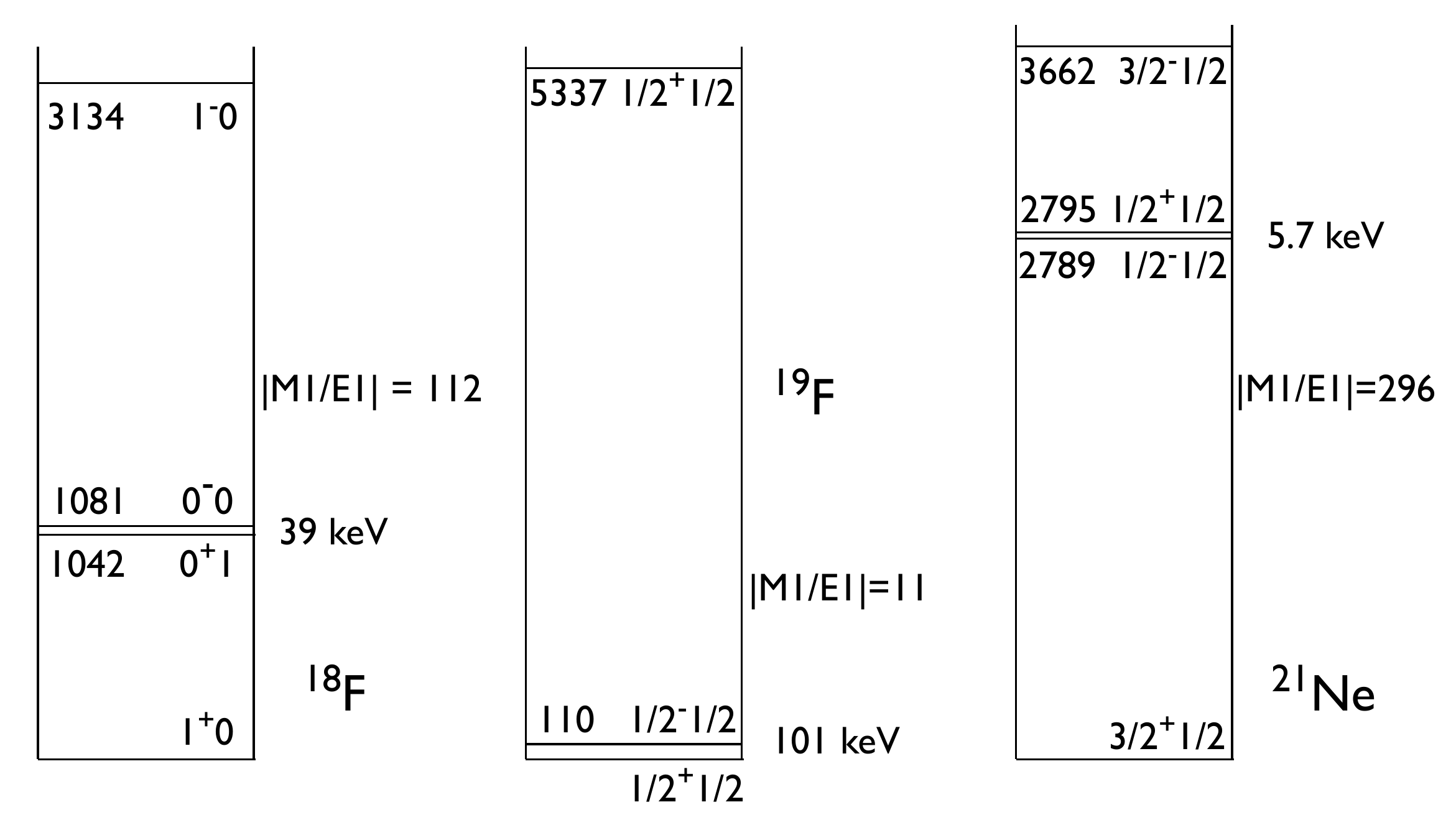}
\caption{Three sd-shell two-level nuclear systems in which PNC observables are enhanced.}
\label{fig:PNCnuclei}
\end{figure}

Figure \ref{fig:PNCnuclei} shows three nuclei of interest, the parity doublets
in $^{18}$F, $^{19}$F, and $^{21}$Ne.  These are effectively two-level systems because
the small parity-doublet splittings (39, 101, and 5.7 keV) make doublet mixing much more
important than mixing with distant states.   The theoretical challenge is to 
identify methods for calculating two-level mixing accurately.

$^{18}$F is an interesting case for illustrating both the sources of PNC enhancement and
the nuclear structure analysis.  The circular polarization of the $\gamma$ ray emitted in the decay
of the 1081 keV $0^-0$ to the $1^+0$ ground state is given by
\begin{equation}
P_\gamma(1081~\mathrm{keV}) = 2~ \mathrm{Re} \left[ { \langle 0^+1| V_{PNC}|0^-0 \rangle 
 \over 39 \mathrm{keV}} { \langle 1^+0 (g.s.) | M1 | 0^+1 \rangle \over \langle 1^+0 (g.s.) | E1 | 0^-0 \rangle} \right]
 \label{eq:18}
 \end{equation}
 As the typical scale of PNC nuclear mixing matrix elements is $\sim$ 1eV, the first ratio
in Eq. (\ref{eq:18})  is  $\sim 10^{-5}$.  The second term is the ratio of a PNC M1 transition
to the normal PC E1 transition.  Both transition strengths are known
experimentally.  The E1 transition is quite suppressed: the leading-order operator vanishes in a self-conjugate nucleus. (It corresponds to a translation
of the center-of-mass.)  The M1 is exceptionally strong, $\sim$ 10.3 W.u.  Thus the second 
ratio is $\sim$ 110.  One concludes that
the expected size of $P_\gamma$ is $\sim 10^{-3}$, four orders of magnitude above the
typical scale of PNC in the NN system.  Everything is known in Eq. (\ref{eq:18})
except the sign of the $M1/E1$ ratio and the mixing matrix element.

Following early work by Barnes et al. \cite{barnes}, heroic 
efforts to measure $P_\gamma$ were made by the
Queens \cite{queens} and Florence \cite{florence} 
groups, yielding $(8 \pm 39) \times 10^{-5}$.  The DDH
best-value prediction is $(208 \pm 49) \times 10^{-5}$.  As the mixing is purely isovector,
the expected enhancement due to neutral currents was not found.

First-principles calculations of PNC mixing matrix elements must address several
difficulties.  The underlying operator is dipole-like $\sim \vec{\sigma} \cdot \vec{p}$ and thus
sensitive to spurious components, so that
projection of the center-of-mass is important.  As this operator couples opposite-parity shells,
configurations in any included space are linked directly to the excluded space,
leading to a sawtooth oscillation of the matrix element as new shells
are added.  The operator behaves under time reversal like the E1 operator, which is
suppressed by correlations.  $V_{PNC}$ is a surface operator, sensitive to the
shapes of the single-particle wave functions.  Most important, the down-side of exploiting
parity doublets is the need to calculate a highly exclusive matrix element, one that
exhausts a tiny fraction of the sum rule generated when $V_{PNC}$ operates on either member
of the doublet. 

In $^{18}$F these difficulties can be avoided by a simple trick:  the doublet PNC mixing
is identical, up to isospin rotation, to the exchange-current contribution to the axial-charge
$\beta$ decay transition between the $0^+1$ ground state of $^{18}$Ne 
(the analog of the $0^+1$ doublet state) and the $0^-0$
member of the doublet.  Furthermore the ratio of the exchange current contribution to
the one-body operator $\vec{\sigma} \cdot \vec{p} ~\tau_-$ is $\sim$ 1 (both operators are
of order $v/c$) and stable: the exchange current is effectively a renormalization of the one
body operator.  One can use the measured $\beta$ decay rate to 
determine the PNC mixing matrix element \cite{haxton}.   This argument, applied in a variety of 
nuclear structure calculations, leads to predictions of  $\langle V_{PNC} \rangle$ that are
stable to about $\pm 7\%$.  The case of $^{19}$F is similar, though there are
additional uncertainties in this case because the mixing matrix element also contains an
isoscalar piece.

\begin{figure}
\includegraphics[width=13cm]{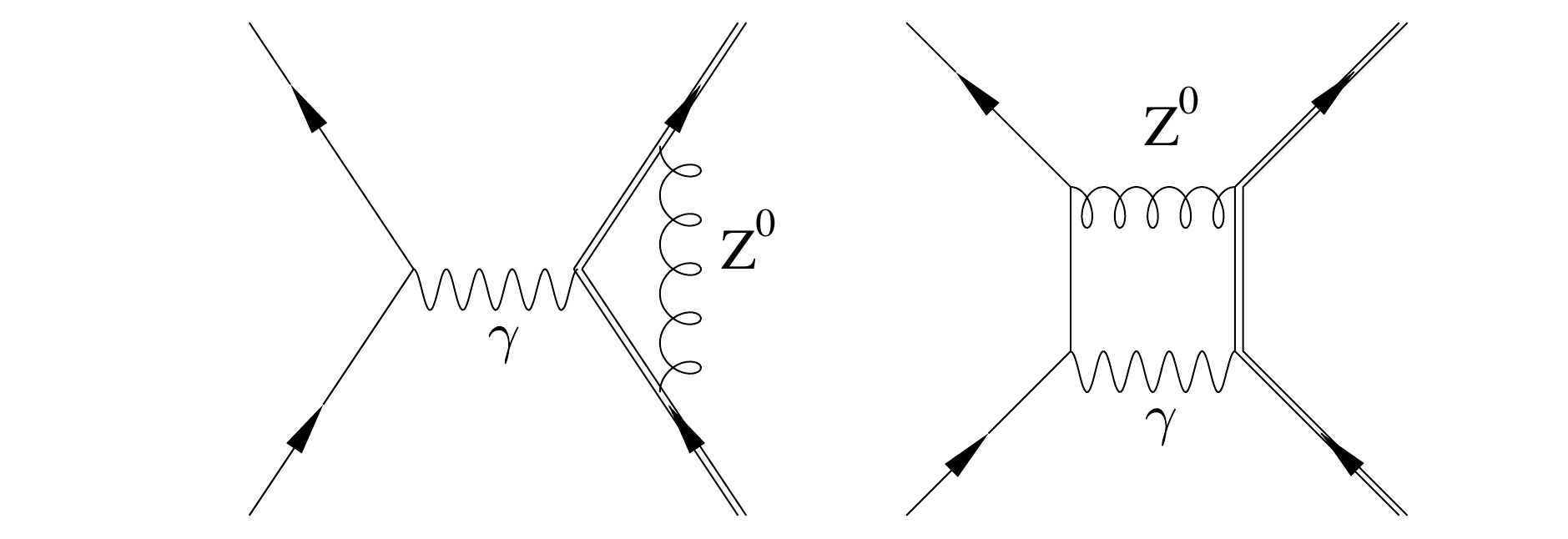}
\caption{Weak radiative corrections contributing to electron-nucleus interactions
include the a) the anapole moment as well as b) terms that do not involve single 
photon exchange.}
\label{fig:anapole}
\end{figure}

Another constraint \cite{anapole,anapole2} comes from atomic PNC experiments in which
the nuclear anapole moment generates a dependence on the nuclear spin.
The anapole moment is a weak radiative
correction to the electron-nucleus interaction (see Fig. \ref{fig:anapole}) that
acts like a contact interaction and grows as $A^{2/3}$.  In a heavy nucleus it can dominate
the tree-level spin-dependent interaction from V(e)-A(nucleus) Z exchange. 
There are various contributions to the anapole moment, but the
most important term comes from nuclear ground-state polarization due to
$V_{PNC}$.  As the case of interest, $^{133}$Cs \cite{wieman}, has no ground-state
parity doublet, the polarization is dominated by mixing with the collective giant resonance
region.

\begin{figure}
\includegraphics[width=15cm]{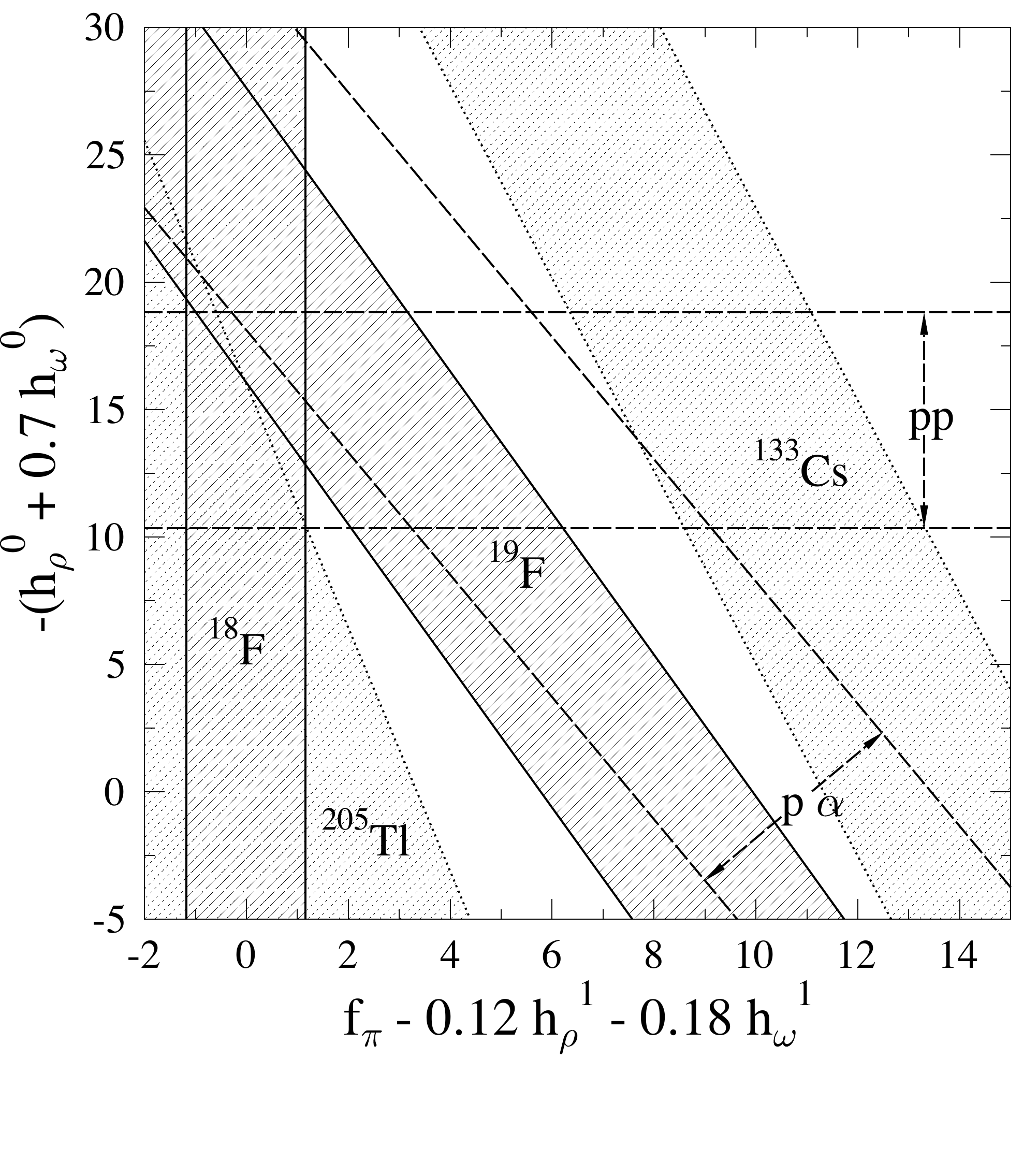}
\caption{Experimental PNC constraints as a function of the effective isoscalar and
isovector couplings.}
\label{fig:results}
\end{figure}

A summary of what we have learned from experiment and theory is shown in 
Fig. \ref{fig:results}.  To a good approximation the observables measured to date depend
on two sets of couplings, one isoscalar $\sim -h_\rho^0-0.7 h_\omega^0$ and
one isovector $\sim f_\pi -0.12 h_\rho^1-0.18 h_\omega^1$.  (The constraint
from $A_L^{\vec{p}+p}$ is plotted assuming the DDH value for $h_\rho^2$.)   The
overall consistency of the results is not high.  It appears from $P_\gamma$ in $^{18}$F
that the isovector coupling, where one expects the neutral current to dominate, is
significantly smaller than the DDH best value.   There is reasonable
agreement between the odd-proton cases, $^{19}$F and $\vec{p}+\alpha$, which intersect
with the $^{18}$F band at a point roughly consistent with the DDH best value for the
isoscalar coupling.  However, the odd-proton constraint from $^{133}$Cs favors larger
values of the couplings.   (A bit of a very broad band from the upper bound on the
$^{205}$Tl anapole moment is also shown: the uncertainty in this result is such that it 
does not impact the conclusions drawn from the Fig. \ref{fig:results}.)

\section{Summary}
The study of hadronic PNC has proven to be a very challenging area: both the experiments
and the theoretical analysis are difficult.  While some
reasonable consistency exists between the $^{18}$F, $^{19}$F, $\vec{p}+\alpha$, and 
$\vec{p}+p$ results (assuming $h_\rho^2$ is near the DDH best value), error bars are large
and there is no significant redundancy among the measurements. 

The conclusion from $^{18}$F that the isovector coupling is small, compared to the DDH
best value,  may be one of the more solid results.  Several $^{18}$F
experiments have placed tight upper bounds on $P_\gamma$, and the analysis,
though it involves a complex nucleus, is unusually free of structure uncertainties.
Consequently,  we have yet
to find evidence for neutral currents in $\Delta$S=0 interactions.  Such suppression,
relative to the isovector strength, is superficially reminiscent of the enhancements
embodied in the $\Delta$I=1/2 rule
in flavor-changing reactions.

Clearly a lot remains to be done.  The ongoing effort
to measure $d\phi/dz$ in $\vec{n}+\alpha$ is important, as the comparison with $\vec{p}+^4$He would
allow an alternative $\Delta$I=0/$\Delta$I=1 separation to be made, checking
the pattern seen in Fig. \ref{fig:results}.   The discrepancy involving
the $^{133}$Cs anapole moment is troublesome.  As the control of systematics in that experiment
required years of effort, it is not clear when the next anapole moment measurement
will be made.  But, from a theoretical perspective, such a measurement in an 
odd-neutron system would be useful in a PNC isospin analysis.  If one
could tighten the constraints on the isoscalar and isovector couplings, $A_L^{\vec{p}+p}$
would become an independent test of $h_\rho^2$.

While progress has been slow over the past decade, new facilities such as the SNS (with
its high-intensity cold neutron beam) and FRIB (a possible source of radioactive nuclei
with enhanced anapole moments) may help the field along in the next few years.

Good progress has been made in theory, with the development of a more systematic
expansion for the effective PNC interaction being one recent example.  But the lack of 
redundancy among experiments puts a lot of stress on theory, requiring one to make
use of constraints in NN, few-body, and nuclear systems.   It is not clear whether
results from NN and few-body systems should be compared naively with those from complex nuclei.
Couplings extracted from complex nuclei are necessarily effective,
defined in the context of chosen shell-model spaces.    We have many examples in
nuclear physics (the axial-vector coupling $g_A$ being a celebrated one) where
the shell-model coupling is not the underlying bare coupling. 
There has been some work comparing PNC
calculations in small included spaces with those in larger ones.  There is a significant
dependence on the model space, indicating that effective couplings may differ substantially
from the bare values.    One good exercise 
for theorists might be to explore this question in a light nucleus where many shells
can be included, in order to test the evolution of $\langle V_{PNC} \rangle$ with shell number.
This could provide some guidance in interpreting results from systems like $^{18}$F.

\section*{Acknowledgments}
I thank C.-P. Liu for helpful discussions.  This work was supported in part by
the U.S. Department of Energy, Office of Nuclear Physics.


\begin{thebibliography}{9}
\bibitem{adelberger} E.~G. Adelberger and W.~C. Haxton, {\em Ann. Rev. Nucl. Part. Sci.} {\bf 35},
        501 (1985).
        
\bibitem{cp} C.-P. Liu, nucl-th/0703008/, to appear in {\em Proc. Second Meeting of the APS
        Topical Group On Hadronic Physics}.
        
\bibitem{danilov} G.~S. Danilov, {\em Phys. Lett. B} {\bf 35}, 579 (1971) and {\em Phys. Lett.} {\bf 18},
        40 (1965). 
        
\bibitem{ddh} B. Desplanques, J.~F. Donoghue, and B.~R. Holstein, {\em Ann. Phys.} {\bf 124},
        449 (1980).
        
\bibitem{dz} V.~M. Dubovik and S.~V. Zenkin, {\em Ann. Phys.} {\bf 172}, 100 (1986).

\bibitem{fcdh} G.~B. Feldman, G.~A. Crawford, J. Dubach, and B.~R. Holstein, {\em Phys. Rev. C}
         {\bf 43}, 863 (1991).
         
\bibitem{km} N. Kaiser and U.~G. Meissner, {\em Nucl. Phys. A} {\bf 489}, 671 (1988) and 
         {\bf 499}, 699 (1989).
         
\bibitem{zhu} S.~L. Zhu, C.~M. Maekawa, B.~R. Holstein, M.~J. Ramsey-Musolf, and U.
         van Kolck, {\em Nucl. Phys. A} {\bf 748}, 435 (2005).
         
\bibitem{cpl} C.-P. Liu, {\em Phys. Rev. C} {\bf 75}, 065501 (2007).
         
\bibitem{bonnpp} P.~D. Eversheim {\it et al.}, {\em Phys. Lett. B} {\bf 256}, 11 (1991).

\bibitem{sinpp} S. Kistryn {\it et al.}, {\em Phys. Rev. Lett.} {\bf 58}, 1616 (1987).

\bibitem{pnp} V. A. Knyazkov {\it et al.}, {\em Nucl. Phys. A} {\bf 197}, 241 (1972).

\bibitem{nprad} J.~F. Cavaignac, B. Vignon, and R. Wilson, {\em Phys. Lett. B} {\bf 67}, 148 (1977).

\bibitem{markoff} D. M. Markoff, {\em J. Res. Natl. Inst. Stand. Technol.} {\bf 110}, 209 (2005).

\bibitem{hespin} J. Lang {\it et al.}, {\em Phys. Rev. Lett.} {\bf 54}, 170 (1985).

\bibitem{mich} A. Micherdzinska, http://www.int.washington.edu/talks/WorkShops/, 2007.
           
\bibitem{barnes} C.~A. Barnes {\it et al.}, {\em Phys. Rev. Lett.} {\bf 40}, 840 (1978).

\bibitem{queens} S.~A. Page {\it et al.}, {\em Phys. Rev. C} {\bf 35}, 1119 (1987) and 
           {\em Phys. Rev. Lett.} {\bf 55}, 791 (1985).

\bibitem{florence} M. Bini {\it et al.}, {\em Phys. Rev. Lett.} {\bf 55}, 795 (1985).

\bibitem{haxton} W.~C. Haxton, {\em Phys. Rev. Lett.} {\bf 46}, 698 (1981).

\bibitem{anapole} W.~C. Haxton, C.~P. Liu, and M.~J. Ramsey-Musolf, {\em Phys. Rev. C}
           {\bf 65}, 045502 (2002) and {\em Phys. Rev. Lett.} {\bf 86}, 5247 (2001).
           
\bibitem{anapole2} V.~V. Flambaum and D.~W. Murray, {\em Phys. Rev. C} {\bf 56}, 1641 (1997).

\bibitem{wieman} C.~S. Wood {\it et al.}, {\em Science} {\bf 275}, 1759 (1997).

\end{thebibliography}
\end{document}